\documentclass[conference]{IEEEtran}
\IEEEoverridecommandlockouts
\usepackage{cite}
\usepackage{amsmath,amssymb,amsfonts}
\usepackage{algorithmic}
\usepackage{graphicx}
\usepackage{textcomp}
\usepackage{xcolor}
\usepackage{verbatim}
\usepackage{url}
\usepackage[htt]{hyphenat}

\def\BibTeX{{\rm B\kern-.05em{\sc i\kern-.025em b}\kern-.08em
    T\kern-.1667em\lower.7ex\hbox{E}\kern-.125emX}}
\begin{document}

\title{On Using Linux Kernel Huge Pages with FLASH, an Astrophysical Simulation Code.\\
}

\author{\IEEEauthorblockN{Alan C.\ Calder, Catherine Feldman, Eva Siegmann, John Dey, Anthony Curtis,
Smeet Chheda, Robert J.\ Harrison}
\IEEEauthorblockA{\textit{Institute for Advanced Computational Science} \\
\textit{Stony Brook University}\\
Stony Brook, United States \\
\{alan.calder, catherine.feldman, eva.siegmann, john.dey, anthony.curtis,
smeetdinesh.chheda, robert.harrison\}@stonybrook.edu}
}

\maketitle

\begin{abstract}
We present efforts at improving the performance of FLASH, a multi-scale, 
multi-physics 
simulation code principally for astrophysical applications, by using huge pages
on Ookami, an HPE Apollo 80 A64FX platform. FLASH is written principally
in modern Fortran and makes use of the PARAMESH library to manage a block-structured
adaptive mesh. We explored options for enabling the use of huge pages with several
compilers, but we were only able to successfully use huge pages when compiling with 
the Fujitsu compiler. The use of huge pages substantially reduced the number of 
translation lookaside buffer misses, but overall performance gains were marginal. 
\end{abstract}


\section{Introduction}

\subsection{Ookami}

Ookami is a development platform featuring Fujitsu A64FX processors that is
supported by the United States National Science Foundation and hosted
by Stony Brook University with additional support from the University 
at Buffalo~\cite{ookamiurl}. The project is meant to provide open access to this
new hardware, and beginning in October, 2022,
Ookami will be an XSEDE level 2 service provider~\cite{ookamixsede}.

Ookami, an HPE Apollo 80, has 174 A64FX Fujitsu compute nodes, each with 32GB high-bandwidth
memory (HBM) and a 512GB SSD for the OS.
Ookami's 700 series A64FX processors consist of four core memory
groups each with 12 cores, 64KB L1 cache, and 8MB L2 cache shared
between the cores and run at 1.8 GHz. These processors use the ARMv8.2--A
Scalable Vector Extension (SVE) SIMD instruction set with a 512 bit vector
implementation, allowing for vector lengths anywhere from 128--2048 bits
(in 128 bit increments)
and enabling vector length agnostic programming~\cite{armv8.2}. Ookami has an 
Infiniband HDR100 fat tree interconnect with 200 gigabit switches,
and a high-performance Lustre file system provides about 800 TB storage.

The expectation is that the A64FX processors will provide high 
performance and reliability for applications with common programming 
models, particularly memory-intensive applications, while
maintaining a good performance to power ratio. Ookami
allows users with a variety of applications
to explore the A64FX architecture, and various codes are
being ported to and performance tested on the machine including our 
application code,
FLASH (described below) \cite{pearc_experiences_2021}. 

\subsection{Thermonuclear Supernovae}

Our science application is in nuclear astrophysics, specifically
thermonuclear (Type Ia) supernovae. These events are thought to occur when 
a compact star known as a white dwarf composed principally of carbon and 
oxygen explodes via a thermonuclear runaway, but are incompletely understood 
despite their widely-accepted use as distance indicators for cosmological 
studies~\cite{JacobyEtAl92, Kirshner09}. Our goal is to model these events 
to explore systematic effects on the brightness~\cite{calderetal2017}, and 
our present study explores one subclass of dim events,
Type Iax supernovae, that are thought to be produced by a pure deflagration 
(subsonic burning front) occurring in a special ``hybrid" white 
dwarf~\cite{Foley_2013,Kromer_2015}.

Thermonuclear supernovae are challenging to model because in addition to including a host
of physics, e.g.\ hydrodynamics, self-gravity, and the  material equation 
of state (EOS) for the degenerate electron/positron plasma of the core of 
the white dwarf star, models must address the vast 
disparity of the scales between the white dwarf radius 
($\sim 10^9\ensuremath{\;}{\ensuremath{\mathrm{cm}}}$) and
the width of the laminar nuclear flame at high densities
($< 1\ensuremath{\;}{\ensuremath{\mathrm{cm}}}$). We employ adaptive mesh 
refinement (AMR), but even with many orders of AMR it is not
possible to resolve the physical flame in a whole-star simulation. 
Thus we also employ a model flame to capture the burning on 
sub-grid scales.

\subsection{The FLASH Code}
The FLASH code is a scientific simulation
software package for addressing
multi-scale, multi-physics 
applications~\cite{Fryxetal00,flash_development}. 
FLASH was originally developed at the Flash Center 
at the University of Chicago, supported by the Department of Energy's Accelerated Strategic Computing 
Initiative (ASCI)~\cite{asci2000} 
to address astrophysical problems involving
thermonuclear flashes, stellar explosions powered by a thermonuclear
runaway occurring on the surface or in the interiors of compact
stars. FLASH continues to be developed for astrophysics~\cite{townselyetal2019},
high-energy-density physics~\cite{flashhed}, and more general problems as FLASH-X,
a new code derived from FLASH with a completely new infrastructure, being funded
by the U.S. Department of Energy's Exascale Computing Project \cite{Oneal2018}.

As part of ASCI, the Flash Center was
allowed access to unclassified versions or partitions of the state of the art 
supercomputers installed at the US National Laboratories~\cite{rosneretal2000}, establishing a history of use on advanced platforms.
FLASH won the SC2000 Gordon Bell Prize, Special Category, for AMR simulations of
reactive flow that achieved 238 GFlops on 6420 processors of ASCI Red at the
Los Alamos National Laboratory~\cite{calder.curtis.ea:high-performance}, simulated weakly compressible 
stirred-turbulence at an unprecedented resolution on the
IBM BG/L machine commissioned at the Lawrence Livermore National Laboratory 
in 2005~\cite{Antypas06,terra.turb}, and
served as a formal acceptance test for both Intrepid, an IBM BG/P,
and MIRA, an IBM BG/Q machine, at the Argonne Leadership Computing 
Facility~\cite{flash_pragmatic}.

FLASH uses AMR to address problems with a wide range of 
physical and temporal scales. The current release of FLASH (version 4.6.2) 
implements AMR using the PARAMESH library~\cite{macneice.olson.ea:paramesh, 
macneice.olson.ea:paramesh*1},
which uses a block-structured mesh. FLASH is parallelized primarily
through MPI, although some solvers have been modified to take advantage of threaded
approaches to parallelization~\cite{flash_pragmatic} and
development continues toward a more general design for better 
allowing threading~\cite{CHIUW2019,RADR}.

The PARAMESH library manages a block structured adaptive mesh, with
the data typically in $16 \times 16 \times 16$ zone blocks 
($16 \times 16$ in 2-d).
The principal data container \texttt{unk} stores variables
such as density,
temperature, pressure, etc., in the form
\texttt{unk(nvar, il\_bnd, iu\_bnd, jl\_bnd, ju\_bnd, kl\_bnd, ku\_bnd, maxblocks)}
where \texttt{nvar} is the number variables, 
\texttt{il\_bnd:iu\_bnd, jl\_bnd:ju\_bnd, kl\_bnd:ku\_bnd} are 
the x, y, and z zone limits, 
and \texttt{maxblocks} is the number of blocks. PARAMESH
is thus designed for loops using data from blocks,
and there is a stride in memory for addressing variables in different zones or
blocks. This feature motivated our interest in investigating the use of
huge pages as a way of improving performance. 

The modified version of FLASH we use for supernova simulations
utilizes an advection-diffusion-reaction (ADR) scheme~\cite{VladWeirRyzh06} 
that propagates reaction progress variables to model the stages of the 
sub-grid-scale flame. Flame speeds are from the tabulated results
of previous calculations~\cite{timmes92,Chamulak2007The-Laminar-Fla} and 
also include enhancement to the burning rate from unresolved buoyancy and 
background turbulence~\cite{Khok95,townsley.calder.ea:flame,jacketal2014}. 
Finally we note that the FLASH simulations use double precision
arithmetic.

\section{Porting and Profiling FLASH}
Our efforts at porting FLASH to Ookami began with identifying the prerequisites
for running our supernova application, e.g.\ compilers, MPI, and HDF5, and 
ensuring these were available. We initially explored 
the GCC 10.3.0, Cray 10.0.3, and Arm 21.0 compilers with  MVAPICH 2.3.5
and OpenMPI 4.0.5. This effort showed that FLASH ran ``right out of the box" with
these and scaled reasonably well with no tuning. 
As different combinations of compilers and MPI implementations have been shown to produce large performance differences for the same code, we were interested to see if this trend persisted with our supernova problem.
We found that the ARM compiler produced an executable that ran almost 2.5 times slower than those created with the Cray and GCC compilers; 
the runtime differences between the latter were negligible. 
However, the same executable compiled using GCC 10.2.0 with MVAPICH 2.3.5 on Intel Xeon E5-2683v3 CPUs ran three times quicker as the fastest runs on Ookami. 
It is important to note that this runtime comparison was done
before tuning our code to use the A64FX's main features, including use of SVE instructions 
and HBM, so accordingly this is where we turned to next.

To see where our code could most benefit from use of these features, we began our investigation into the performance of FLASH on Ookami
by profiling with Arm MAP~\cite{ArmMap}. Our scientific
investigation of Type Iax supernova is currently performing
suites of 2-d simulations that allow for a relatively inexpensive 
exploration of both the Ookami platform and the parameter space of 
the astrophysics problem. Our MAP study
indicated that FLASH spent considerable time in the routines for
the EOS. 
Accordingly, we decided to 
analyze that part of the code while running 2-d supernova simulations. 

The next step was to investigate use of SVE, but taking advantage of SVE 
proved difficult and
we found that considerable modification will be required to vectorize our
version of FLASH. The problem is the vast scope and branching of the
main loops in the EOS routines that prevents using SVE.

Accordingly, in an effort to identify other issues to address
to boost performance we instrumented the code with the Performance Application Programming Interface (PAPI)~\cite{papi}, and we describe this process in detail below. 
The finding of that analysis was that the number of data translation lookaside buffer (DTLB) misses were exceptionally high, so we focused on the use of huge pages in
memory as the next avenue of investigation, which is the main topic of the paper. 
A full report on our exploration with scaling results and details of our 
attempt to utilize the A64FX's SVE instructions and NUMA architecture may be
found in~\cite{flashexperience2022}.

The 2-d supernova problem with the EOS was the obvious choice for our
 investigation into the use of huge pages.
Eventually, however, we will run full 3-d simulations of supernovae, so to 
test the principal
component of 3-d simulations, the hydrodynamics routines, we settled on 
also investigating a pure
hydrodynamics simulation, the Sedov explosion problem~\cite{sedov1959},
one of the standard test problems provided with FLASH. 

For the two problems,
we instrumented the code to record the performance of the routines 
of interest, the EOS routines in the supernova case and the hydrodyamics
routines in the Sedov case. 
Thus in this work we report on two tests 
dubbed ``EOS'' and ``3-d Hydro'' because
those were the parts of the code we instrumented for performance
measurements. Details of both the
EOS and the Sedov explosion problem
may be found in the original FLASH paper~\cite{Fryxetal00}. The EOS test
ran a 2-d supernova simulation for 50 time steps and the 3-d Hydro test
ran a Sedov explosion simulation for 200 time steps.

Our instrumentation of FLASH with PAPI began by identifying a subset of PAPI events that can characterize 
overall performance --- use of SVE measured as SVE instructions 
per cycle, memory bandwidth, DTLB misses, and the 
number of hardware cycles, and we tested 
the use of PAPI to monitor these quantities with several simple 
example programs. 

Following the Object Oriented Programming in Fortran 2003 lesson,
``Tutorial OOP(III): Constructors and Destructors" by Danny Vanpoucke,
we began instrumenting FLASH by constructing a Fortran
object to interface with the PAPI routines~\cite{vanpouckeOOP}. 
We used a Fortran module to initialize the object and allocate
member variables, call the PAPI begin function, and finalize by calling
the PAPI end function and deallocate the variables.
The object also used another Fortran module to store an identifier for the region of the code that was 
instrumented. Using the Fortran \texttt{block} construct allows
avoiding the Fortran requirement that declarations be made before
any executable statements and instantiating the performance
object at any point in a routine.


This module worked with the GNU compiler version 11.2.0 and worked
(with a slight modification) with the Cray compiler version 10.0.3
(both SVE and non-SVE versions). Unfortunately, however, this module
did not work with the Fujitsu compiler version 4.5. The issue was
with calling the finalizer, and we attributed the problem to
the compiler having not reliably implemented
initializers and finalizers of the Fortran 
standard~\cite{fortranbug1,fortranbug2}. So we fell back to 
just ``hard coding" the PAPI calls for the quantities described above
to work with all compilers we tested. 

FLASH also has internal timers, and we noted those results as
a consistency check on our PAPI instrumentation. The timers
record the elapsed time for the simulation, and we
record these
below in addition to the PAPI results. 
We also report the SVE results from PAPI although the code 
was not vectorized
and these results are not critical for this study. At some point 
recalling these ``un-tuned" results may useful. 

\section{Testing Use of huge pages}

At the time we performed our investigation, the operating 
system on Ookami was CentOS 8.1 with kernel 4.18.0-147.el8.aarch64,
which allowed huge pages for memory access. On its own, the kernel
should invoke transparent huge pages, 
an abstraction layer that automates most aspects of creating, managing, and using huge pages~\cite{THP},
when it processes a file greater than 2 Gb. 
In this case, the kernel does most of
the work, though one can alter that with 
the \texttt{hugectl} utility of the \texttt{hugetlbfs} filesystem of Linux~\cite{hugetlbfs} with commands of the form
\texttt{hugectl --shm --thp \ldots} and \texttt{LD\_PRELOAD=libhugetlbfs.so \ldots}
for any compiler and with the 
\texttt{XOS\_MMM\_L\_HPAGE\_TYPE} environment variable 
for the Fujitsu compiler. While the documentation mentions that acceptable values are \texttt{none} or \texttt{hugetlbfs}, ~\cite{fugaku_codesign_report} mentions another possible value \texttt{thp} for the variable on Fugaku and is accepted by the FX700 processor as well.


Huge pages can also be invoked via mechanisms
such as the \texttt{hugeadm} tool of the \texttt{libhugetlbfs} library~\cite{libhugetlbfs_github} and as part of our
exploration we configured two nodes of Ookami
for using these tools (described below).

%
%

We monitored the use of huge pages by looking at system variables 
in \texttt{/proc/meminfo}
that would have values if the huge pages were in use,
\texttt{AnonHugePages}, \texttt{ShmemHugePages}, \texttt{HugePages\_Total},
\texttt{HugePages\_Free}, \texttt{HugePages\_Rsvd}, \texttt{HugePages\_Surp}, 
\texttt{Hugepagesize}, \texttt{Hugetlb}.
Our tests consisted of running the instrumented code with and without
huge pages, while monitoring the values of the variables
in \texttt{/proc/meminfo} to ensure that huge pages were in use when
expected. 

Our investigation into huge pages also involved modifications to 
Ookami. We installed the Fujitsu compiler and also modified the
system on two Ookami nodes~\cite{hugelibrary,hugenode1,hugenode2}.  
The modifications to the two nodes include:

\begin{itemize}
%
\item Installation of the Fujitsu compiler, which required
kernel boot time parameters to be set.

\begin{itemize}
\item Kernel boot parameters set: 
\texttt{hugepagesz=2M hugepagesz=512M default\_hugepagesz=2M}

\item Setting kernel parameter 
in \texttt{/etc/sysctl.d/98-fujitsucompilersettings.conf} 
\texttt{kernel.perf\_event\_paranoid=1}

\end{itemize}

\item Installation of the package \texttt{libhugetlbfs-utils}, specifically
the tool: \texttt{hugeadm} for configuring the hugepage environment.

\begin{itemize}

\item Specialized UNIX group created for empowered end user 
use: 'hugetlb\_shm\_group' and added members.

\item Allowed enabling and disabling transparent huge pages by changing 
settings in the file \texttt{/sys/kernel/mm/transparent\_huge\-page/enabled}.

\begin{itemize}	

\item Enable transparent huge pages by setting the file contents to “[always]
	madvise never” (command: \texttt{echo always > /sys/kernel/mm/transparent\_huge\-page/enabled}.

\item Disable transparent huge pages by setting the file contents to “always 
	madvise [never]” (command: \texttt{echo never > /sys/kernel/mm/transparent\_hugepage/enabled}.

\end{itemize}

\end{itemize}

\end{itemize}

\section{FLASH Results}

Our first test was to compile FLASH with the GNU compiler
to demonstrate that it was able to use huge pages. Try as we might, 
however, we were not able to use huge pages when FLASH was compiled 
with the GNU compiler. We tried many variations of explicitly using
the \texttt{hugectl} tools, explictly linking the \texttt{hugetlbfs} library, and
using the modified nodes, all to no avail.

To further investigate, we wrote two simple Fortran test programs, 
one statically allocating memory for a 2-d array and one dynamically 
allocating memory for a 2-d array, and then just repeated calculating sums
over the arrays. As expected, the program with the dynamically allocated 
array was able to use huge pages with the GNU compiler while the statically 
allocated array version could not. This behavior is expected because
transparent huge pages only maps anonymous memory regions 
such as heap and stack space \cite{THP}. 

In normal usage, FLASH is compiled with specific values of
the block sizes, e.g.\ $16 \times 16 \times 16$, for the variables
in the data container \texttt{unk} and
changing any of those requires re-compiling the code. PARAMESH 
also has a library mode that is designed to allow use of it
as a pre-compiled library so that the block sizes may be set 
at run time. Our interpretation was that FLASH using PARAMESH
with the block sizes set at compile time was statically allocating
the array \texttt{unk}. Investigation showed that this was not
the case, however, and \texttt{unk} was dynamically allocated
even when PARAMESH was not in library mode. Apparently at
some point the FLASH or PARAMESH developers switched to dynamic
allocation to avoid difficulties with statically allocating 
large arrays \cite{library}.

We subsequently found that FLASH also did not use huge pages
with the Cray compiler. Why FLASH does not use huge pages with
the GNU and Cray compilers remains a mystery, unfortunately. 
We did find success with the Fujitsu compiler. In that case, FLASH
naturally used huge pages and use of huge pages had to be explicitly
turned off via the \texttt{-Knolargepage} flag for compiling and linking.  
Accordingly we have a performance comparison only for the Fujitsu compiler.

Tables \ref{supernovatab} and  \ref{sedovtab} present results for the EOS
and 3-d Hydro problem, respectively. 
Shown are results for the five PAPI performance measures and 
the elapsed time from the FLASH timer for simulations run 
with and without huge pages. 

\begin{table}[htbp]
\caption{Results with the Fujitsu Compiler for the EOS problem}
\begin{center}
\begin{tabular}{|l|c|c|}
\hline
	\textbf{Measure} & \textbf{\textit{Without HPs}} & \textbf{\textit{With HPs}} \\
\hline
Hardware (cycles)   & 1.25 $\times 10^{11}$ & 1.17 $\times 10^{11}$ \\
\hline
Time (s)            & 6.97$\times 10^{1}$  & 6.52 $\times 10^{1}$ \\
\hline
SVE Instructions/cycle    & 0.47  & 0.51   \\
\hline
Memory (Gbytes/s)   & 4.19 & 4.45 \\
\hline
DTLB misses (1/s)   & 2.34 $\times 10^{7}$ &  1.10 $\times 10^{6}$ \\
\hline
FLASH Timer (s)     & 339.032 & 333.150 \\
\hline
\end{tabular}
\label{supernovatab}
\end{center}
\end{table}

\begin{table}[htbp]
\caption{Results with the Fujitsu Compiler for the 3-d Hydro problem}
\begin{center}
\begin{tabular}{|l|c|c|}
\hline
	\textbf{Measure} & \textbf{\textit{Without HPs}} & \textbf{\textit{With HPs}} \\
\hline
Hardware (cycles)   & 1.21 $\times 10^{12}$ & 1.20 $\times 10^{12}$ \\
\hline
Time (s)            & 6.70 $\times 10^{2}$ & 6.69 $\times 10^{2}$ \\
\hline
SVE Instructions/cycle  & 0.11  & 0.11   \\
\hline
Memory (Gbytes/s)   & 10.10 & 10.09  \\
\hline
DTLB misses (1/s)   & 2.42 $\times 10^{6}$ & 7.83 $\times 10^{5}$ \\
\hline
FLASH Timer (s)     & 1203.616 & 1176.312 \\
\hline
\end{tabular}
\label{sedovtab}
\end{center}
\end{table}

\begin{figure}[htbp]
\centerline{\includegraphics[width=\columnwidth]{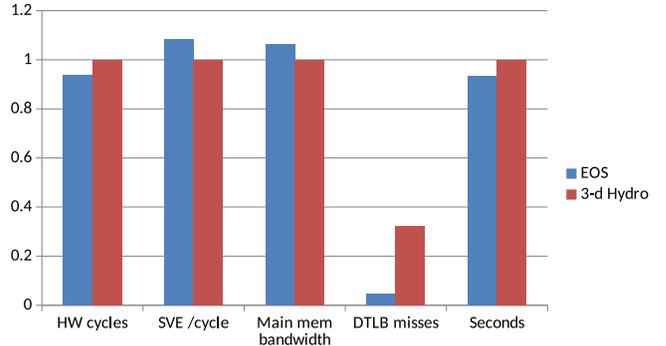}}
\caption{Bar chart showing the ratios of
	PAPI performance measures for simulations with and
	without huge pages. The EOS case is in blue and
	the 3-d Hydro case is in red.}
\label{fig}
\end{figure}

Figure \ref{fig} compares the results with and without use of huge pages 
for the two simulations. Shown is a bar chart with the 
ratio of each performance measure using huge pages to
the measure without use of huge pages for the two test
simulations. All measures
but DTLB misses are close to one, showing little effect when using huge
pages. The low ratios for DTLB misses (0.047 and 0.324 for the EOS and
3-d Hydro tests, respectively) show that use of huge pages drastically
reduces these misses. The near unity ratios for the times, however,
show the reduction of DTLB misses did not significantly improve the
performance for either problem.

\section{Summary and Conclusions}

We found that the FLASH code was only able to use huge pages when 
compiled with the Fujitsu compiler. The reason executables 
compiled with the GNU and Cray compilers did not use huge pages remains 
a mystery. A small test program that dynamically allocated memory 
was able to use huge pages when compiled with the GNU and Cray
compilers (in addition to the Fujitsu compiler) as expected. 
The main data containers in FLASH are dynamically 
allocated, so our first proposed explanation of the 
problem,  static allocation of arrays in FLASH,
does not explain the behavior.

We found that when using huge pages, simulations that exercised
physics modules in FLASH known to have significant computational
expense showed a dramatic reduction in DTLB misses. This reduction,
however, did not significantly improve the performance of these
simulations. Further investigation is needed to understand why.

Installing the Fujitsu compiler on Ookami required installing packages 
for huge pages. While we modified two nodes of Ookami to enable
huge pages, the unmodified nodes also readily used huge pages upon installing 
the Fujitsu compiler and did not demonstrate
different behavior from modified nodes. We suspect the modifications 
were either unnecessary or redundant with the installation of the Fujitsu 
compiler. 


\section*{Acknowledgment}
The authors would like to thank Stony Brook Research Computing and
Cyberinfrastructure, and the Institute for Advanced Computational Science
at Stony Brook University for access to the innovative high-performance
Ookami computing system, which was made possible by a \$5M National
Science Foundation grant (\#1927880).
This research was supported in part by the US Department of Energy (DOE) 
under grant DE-FG02-87ER40317. FLASH was developed in part by the US Department 
of Energy (DOE) NNSA-ASC and OSC-ASCR-supported Flash Center for Computational 
Science at the University of Chicago.
The authors gratefully acknowledge the generous support 
of the Ookami community, the support team and other users,
that made this work possible and a joy. The authors also
thank Klaus Weide for insightful discussions.


\bibliographystyle{IEEEtran}


\end{document}